\documentclass[epj]{svjour}
\usepackage{amsmath}
\usepackage{latexsym}
\usepackage{amssymb}
\usepackage{mathtools}
\usepackage{braket}
\usepackage{graphicx}
\usepackage{dsfont}
\usepackage[colorlinks=true, citecolor=blue, urlcolor=blue]{hyperref}
\usepackage{float}
\usepackage{amsfonts}

\begin{document}
\title{Resource state structure for {controlled} quantum key distribution}

\author{Arpan Das\inst{1,2,}\thanks{\emph{Email:}{arpandas@iopb.res.in}}\and Sumit Nandi\inst{1,2,}\thanks{\emph{Email:}{sumit@iopb.res.in}}\and Sk Sazim\inst{3,}\thanks{\emph{Email:}{sksazim@hri.res.in}}
\and Pankaj Agrawal\inst{1,2,}\thanks{\emph{Email:}{agrawal@iopb.res.in}}}
\institute{Institute of Physics, Sachivalaya Marg, Bhubaneswar 751005, Odisha, India.\and Homi Bhabha National Institute, Training School Complex, Anushakti Nagar, Mumbai 400085, India.\and Harish-Chandra Research Institute, HBNI, Allahabad 211019, India.}

\date{Received: date / Revised version: date}

\abstract{
Quantum entanglement plays a pivotal role in many communication protocols, like secret sharing 
and quantum cryptography. We consider a scenario where more than two parties are involved in a protocol
and share a multipartite entangled state. In particular, we considered the protocol of {Controlled}
Quantum Key Distribution (CoQKD), introduced in the Ref. Chin. Phys. Lett. 20, 183-185 (2003), where, two parties, Alice and
Bob establish a key with the cooperation of other parties. Other parties control/supervise whether Alice and Bob
can establish the key, its security and key rate.  We discuss the case of three parties in detail and find 
suitable resource states. We discuss the controlling power of the third party, Charlie. We also examine the usefulness of the
new resource states for generating conference key and for cooperative teleportation. We find that recently introduced Bell inequalities can be useful to establish the security of the conference key. We also generalize the scenario to more than three parties.}

\maketitle

\section{Introduction}
Quantum entanglement is an exquisite resource behind many   
quantum information processing protocols. In particular, usefulness of entanglement as a resource in cryptography was demonstrated by Ekert \cite{Ekert} in a protocol which was an extension of the seminal BB84 scheme \cite{BB84}. 
Since then, many variants of this protocol have been proposed using bipartite and multipartite entanglement \cite{qkd}. In the Ekert's protocol, Alice and Bob share a set of Bell states and make measurement in three bases each. Of these, two bases are common. There are nine combinations of bases, among which two combinations give correlated results and are used to establish the secret key. Four combinations of bases are used to establish the security using the violation of Bell-CHSH inequality \cite{chsh} and the rest are discarded. We consider a variant of this protocol using multipartite states.

In a multiparty case, we shall consider two scenarios. In the first scenario, a key
is established between two parties, say Alice and Bob, with other parties
controlling this key generation. {This protocol of Controlled QKD (CoQKD) has also been discussed in the Ref. (\cite{cpl})}.
In the second scenario, a secret key is established among
all the parties.
For the first scenario, need for control may
arise for a number of different reasons. Some of them could be: i) one of the two parties may be dishonest, ii) one of the two parties may be compromised by some eavesdropper, iii) the communication may be done only under some supervision.
In this protocol, the controller/supervisor determines the state that Alice
and Bob can share. This state can be a product state, a partially entangled state, or a maximally entangled state. The nature of this state will determine if a key can
be established or not, and the key rate and security. This protocol is different from  MDI-QKD protocols \cite{mdi-qkd1,mdi-qkd2}, where the third party (say Charlie) is an untrusted party. Alice and Bob send two photons to Charlie and he projects the joint state onto a Bell state. Whereas, in Controlled QKD, Charlie's measurement is the first step of the protocol. The collapsed state between Alice and Bob after the measurement of Charlie is used by them to establish the secure key. In the next section, we illustrate various features of this protocol using three-qubit GHZ states. We then find all three-qubit resource states that would allow the generation of a
perfect key with optimal key rate, and then generalize to more than three parties.
We also discuss the security of the protocol using Bell-CHSH inequality.

In the second scenario, the key, that is established among all parties, is known
as conference key. In a conference key, all parties share a common key that they can use for secret communication with one another. Using the new resource state (called NMM-state), introduced for CoQKD, we give a protocol
to establish a conference key. 
 We use our recently introduced Bell inequalities \cite{arpan,arpan1}  to establish the security of this conference key protocol. 

These resource states can be thought of as task-oriented
maximally entangled states (TMESs) \cite{tmes}, where task is {\em maximal} Co-QKD, {\it i.e.} the generation of a perfect secret key with optimal key rate. We also 
show how these states can be generated from product states by using suitable multinary unitary operators. It turns out that these resource states have the structure that makes them suitable for cooperative quantum teleportation protocol also. We discuss some aspects of this protocol using new resource states.

The paper is organized as follows. In Sections \ref{sec0}, \ref{sec1}, \ref{sec2} and \ref{sec3}, we cover the three-qubit scenario in detail. Specifically, in the section \ref{sec0}, we introduce the CoQKD protocol. In the section \ref{sec1}, we propose the structure of three-qubit resource states suitable for maximal CoQKD. Then, we discuss the CoQKD protocol using these resource states in  the section \ref{sec2}. 
In the  section \ref{sec3}, we show the generation of the conference key by these resource states.
In the  section \ref{sec4}, we explore further for the suitable structure of states useful in CoQKD beyond three qubits.  We discuss how these new resource states are also suitable for cooperative teleportation in  the section \ref{sec5}. In the  section \ref{sec6}, we identify these resource states as TMESs. Lastly, we conclude in the  section \ref{sec7}.

\section{CoQKD Protocol }\label{sec0}

In this section, we {illustrate} the CoQKD protocol using three-qubit
GHZ state. The goal of this protocol is to establish a secret key
between two parties with the involvement of the other parties. As discusses
earlier, there could be multiple reasons for the involvement of other
parties. One of the parties with the final secret key is not trustworthy,
 and wants to disrupt the key creation. Specifically, in the key forming procedure,
 the dishonest party can deliberately make false statements in the public declaration 
 rounds and affect the secret key. Moreover, he/she can be compromised by an 
 external attacker, such that they two collaborate to act dishonestly. 
If there is a supervisor who can control the shared entangled state between the two parties and supervises the key generation then a party cannot cheat or be compromised by any external eavesdropper. Specifically, the supervisor knows exactly which state the two parties are sharing and what the optimal key rate should be. If the parties report the controller/supervisor a smaller key rate than the optimal, then there is a cheating involved and that run of the key generation is discarded. The supervisor is also a controller. All the parties share a multipartite entangled state and the controller does a measurement on his/her system to initiate the key generation process. There may be more than one controller and one controller may have more than one qubit in his/her possession.

 In this protocol, there are $N$ ($N\geq 3$) parties, among which $N-2$ parties control the secret key generation by two remaining parties, say Alice and Bob. To illustrate the protocol clearly, we consider a simple scenario of three parties who share a three-qubit GHZ state, $\ket{GHZ}=\frac{1}{\sqrt{2}}(\ket{000}+\ket{111})$. If the controller Charlie performs measurement on his qubit in Hadamard basis, then the collapsed state between Alice and Bob is a Bell state: $\ket{\psi^{\pm}}=\frac{1}{\sqrt{2}}(\ket{00}\pm\ket{11})$. After the measurement, Charlie announces his choice of basis publicly. Then following the Ekert's protocol, Alice and Bob can establish a perfect secret key with optimal key rate between them, {\it i. e.} without any Quantum Bit Error rate (QBER) . 
After establishing the key, Alice and Bob both report the key rate to Charlie. If it is not optimal, then Charlie knows that there has been a cheating. Moreover, without the measurement performed by Charlie, Alice and Bob can not establish a key, as their qubits are in maximally mixed state. 
So, Charlie also supervises the starting of the key generation process, which allows him to control the whole process.
If Charlie makes a measurement in a basis other than the Hadamard basis, shared state between Alice and Bob is not a Bell state, but a partially entangled state. With this state, perfect key generation with optimal key rate would not be obtained. If the key rate reported to Charlie is less than the optimal w.r.t the shared state between Alice and Bob, Charlie knows that there is a cheating involved and that run of key making is discarded.
%
\begin{figure}[h]
\centering
\includegraphics[scale=0.25]{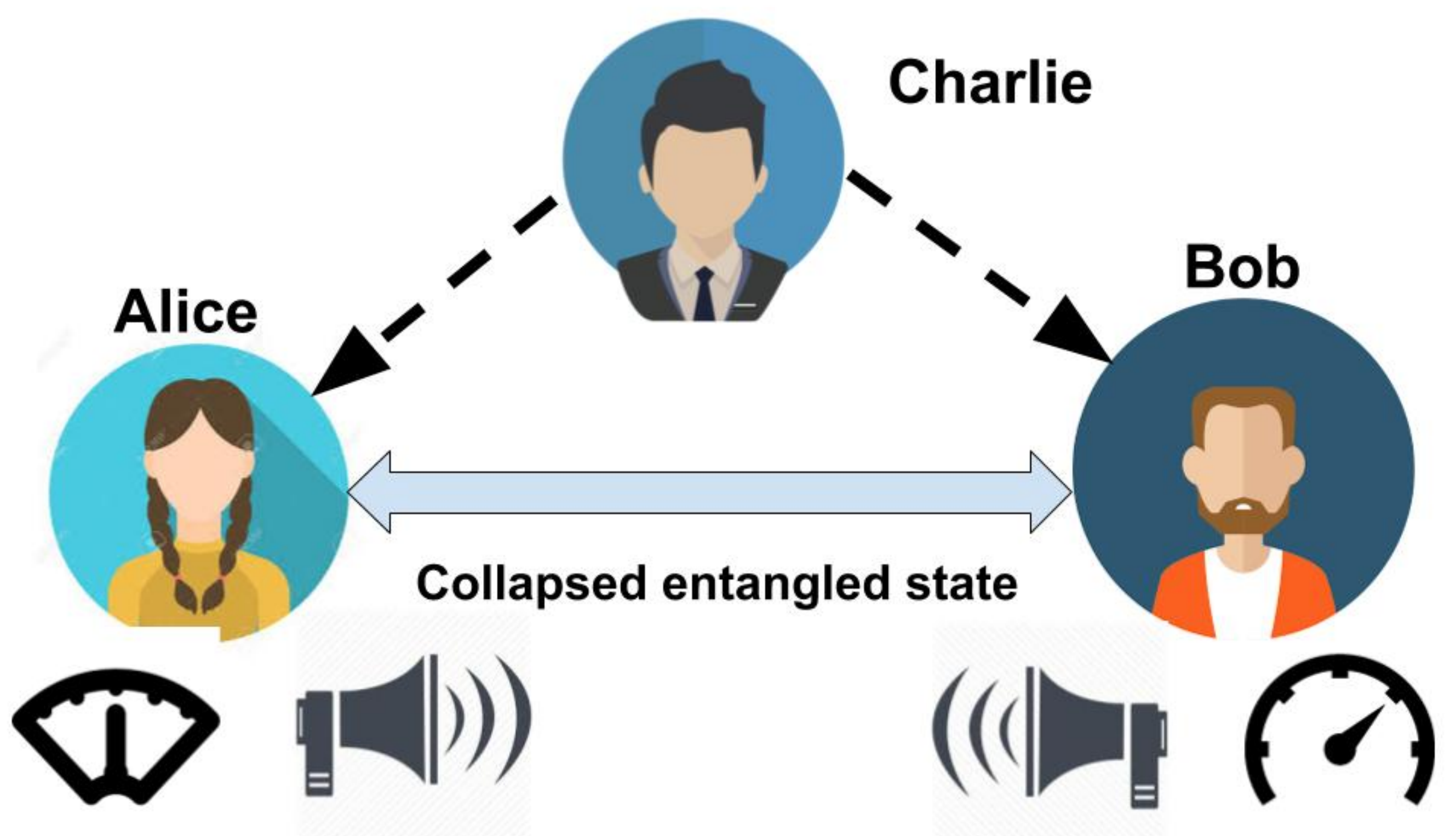}
\caption{CoQKD: Charlie supervises and controls the secret key between Alice and Bob.}
\label{CoQKD}
\end{figure} 
Therefore, for a particular basis of Charlie, Alice and Bob can establish a perfect secret key with optimal key rate, as the collapsed state between Alice and Bob is a maximally entangled state. The key is made according to the Ekert's protocol where the security is certified by the violation of CHSH \cite{chsh} inequality.
If Charlie performs the measurement in a general
basis, the secret key rate is not optimal. In this case we introduce a modified version of Ekert's protocol, where security is guaranteed by the violation of CHSH inequality. Security aspects and 
key rate are discussed in more detail in the section 4.

Advantage of this protocol is that Charlie can detect if there is a cheating involved in the establishment of the secret key and also he can control the optimal key rate and QBER between Alice and Bob. 

In this section, we have seen that for CoQKD, GHZ state can be used to establish a perfect key with
optimal key rate. Therefore,
we say that GHZ state is suitable for {\em maximal} CoQKD. The next question we ask -- are there 
other three-qubit states that may be suitable for {\em maximal} CoQKD? In the case of GHZ state, all
three qubits are maximally mixed. In the next section, we prove that the states, with two maximally mixed marginals, are suitable for {\em maximal} CoQKD.
For a three-qubit pure state, when the first qubit is non-maximally mixed and the other two of the qubits are maximally mixed, we call such a state NMM-state, $\ket{NMM}$. Similarly, there are  MMN or MNM states. These are 
one parameter classes of states. For a specific value of the parameter, these states becomes the GHZ-state. The GHZ state can be thought of as MMM-state, $\ket{MMM}$ because its all marginals are maximally mixed. We show that, apart from the GHZ-state, the other three classes, {\em i.e.} NMM, MNM, and MMN states are also suitable for {\em maximal} CoQKD.

Next, we extend these protocols to four-qubit states. For four qubits, there exist MMNN-states, MMMN-states, or MMMM-states (like GHZ-state, or cluster state) for CoQKD.
Two cases may arise, where each party has one qubit each, or one of the parties
have more than one qubit. We discuss this and the further generalizations in subsequent sections.

\section{Resource state structure for three qubits}\label{sec1}
As discussed above, to establish a perfect secret key with optimal key rate between two parties namely Alice and Bob, 
they must share a Bell state. So, for a controlled scheme what we need is that after all the measurements performed by the controlling parties, the reduced state between last two parties is a Bell state. 
Therefore, the question we ask is: What kind of three-qubit states have this particular feature? Answer to this question is in the proposition below.\\

\textit{\textbf{Proposition-1}} : \textit{The necessary and sufficient condition for a resource state to be suitable for maximal CoQKD protocol is that either all three single qubits  are maximally mixed or only two of 
 the qubits are maximally mixed and the non-maximally mixed qubit is with the controlling party.}\\

\textit{Proof}.-- 
To prove the above statement we proceed in two steps. First we argue that for a successful and maximal CoQKD a specific structure of state is needed and this structure has the entropy distributions as mentioned in the proposition. This is the necessary condition. Secondly, to prove the sufficiency, we show that any state with the above entropy structure is LU equivalent to the specific state useful for maximal CoQKD. The proof is as follows.\\ 
For a successful establishment of a perfect key between Alice and Bob with optimal key rate, they have to share a Bell state. 
That means after Charlie performs the measurement, the collapsed state between Alice and Bob has 
to be local unitary (LU) equivalent to a Bell state. Without loss of generality,
we take that Charlie is performing measurement in the $\{\ket{0}$, $\ket{1}\}$ basis. Then the suitable state for a successful
 controlled QKD is LU equivalent to the following state,
\begin{equation}
\label{threequbit1}
\ket{\Phi}=\sqrt{1/2}\big[\ket{0}\ket{\phi^+}+\ket{1}(\mathbb{I}\otimes U)\ket{\phi^-}\big],
\end{equation}
where, $\ket{\phi^+}=\frac{1}{\sqrt{2}}(\ket{00}+\ket{11})$ and $\ket{\phi^-}=\frac{1}{\sqrt{2}}(\ket{00}-\ket{11})$.
Obviously, for each measurement outcome for Charlie, the reduced state between Alice and Bob is a Bell state or its LU equivalent state. For this state the qubits of Alice and Bob both have entropy one and the qubit of Charlie has entropy less than one if, $U$ is not identity. If $U=\mathbb{I}$, then all the qubits have entropy one and it is just LU equivalent to the conventional GHZ-state. Therefore, the state has the entropy structure as stated in the proposition. To complete the proof, we show that any state with above entropy structure is suitable for maximal CoQKD. To show this we use the results of Refs. \cite{krausprl,krauspra}, which provide details about the local unitary equivalence of multipartite states. Firstly they showed that for three qubits if all the single qubit reduced density matrices are maximally mixed, then they are LU equivalent to $\sqrt{1/2}\big[\ket{0}\ket{\phi^+}+\ket{1}\ket{\phi^-}\big]$, which is again LU equivalent to GHZ-state. This proves our first case, where all the qubits have entropy one. Next case is, when only two qubits have entropy one each. Now, any three-qubit pure state can be written in a Schmidt decomposition between 1-23 bipartition \cite{cohen} as 
\begin{equation}
\label{threequbit2}
\ket{\psi}=\sqrt{p}\ket{0}_C \ket{\psi_0}_{AB}+\sqrt{1-p} \ket{1}_C \ket{\psi_1}_{AB},
\end{equation}
where, $\ket{\psi_0}$ and $\ket{\psi_1}$ both are normalized and $\bra{\psi_0}\psi_1\rangle=0$. We assume that Charlie holds the first qubit; Alice and Bob hold second and third qubit respectively. One can choose different parametrization of these two orthonormal states, such that total number of parameters of the state $\ket{\psi}$ is five. One simple parametrization is the LPS \cite{gisin} scheme. Now we wish this state to have maximally mixed reduced density matrices for $A$ and $B$. This makes the structure to be LU equivalent to,
\begin{equation}
\label{threequbit3}
\ket{NMM}=\sqrt{p}  \ket{0}_C \ket{\phi^+}_{AB} + \sqrt{1-p} \ket{1}_C \ket{\phi^-}_{AB},
\end{equation}
with $p\neq1/2$, as only Bell states and its LU equivalent states have both single qubit reduced density matrix maximally mixed.
The last step is to show that Eq. (\ref{threequbit1}) and Eq. (\ref{threequbit3}) are LU equivalent. We again use the results of the Ref. \cite{krausprl,krauspra}. For two non-generic three-qubit states with two single qubit reduced density matrices maximally mixed are LU equivalent, if the third qubit has same entropy for the both the states. Therefore, Eq. (\ref{threequbit1}) and Eq. (\ref{threequbit3}) are LU equivalent if we can show that for every $p$ we can choose a unitary $U$, such that the qubit with Charlie has the same entropy for both the states in Eq. (\ref{threequbit1}) and Eq. (\ref{threequbit3}). We can take a very simple one parameter unitary,
\begin{equation}
U=
  \begin{bmatrix}
    \sqrt{a} & \sqrt{1-a}  \\
    -\sqrt{1-a} & \sqrt{a} 
  \end{bmatrix}.
\end{equation}
Applying this unitary to Eq. (\ref{threequbit1}) and then tracing out Alice and Bob, we calculate the entropy of reduced density matrix for Charlie and find that for $p=1/2(1-\sqrt{a})$, Charlie's qubit has same entropy for both states Eq.  (\ref{threequbit1}) and Eq. (\ref{threequbit3}) and hence they are LU equivalent. This completes our proof. 
$\square$

\section{Controlled QKD protocol with NMM-state}\label{sec2}
In this section, we discuss controlled QKD using NMM-state. 
Since NMM-state reduces to 
GHZ-state (MMM-state), when $p={1 \over 2}$, our this discussion is more general. 
From Eq. (\ref{threequbit3}), it is evident that if Charlie makes the measurement in $\{\ket{0},\ket{1}\}$ basis then the collapsed state between Alice and Bob is a maximally entangled state. But if Charlie chooses an arbitrary basis then the collapsed state may be anything, {\it i.e.}, separable, partially entangled or maximally entangled. This shows the controlling power of Charlie. Let us consider the arbitrary measurement basis scenario. We take the general basis as,
\begin{equation}
\label{basis}
\ket{+_n}=\frac{\ket{0}+n\ket{1}}{\sqrt{1+|n|^2}},\hspace{2mm} \ket{-_n}=\frac{-n^*\ket{0}+\ket{1}}{\sqrt{1+|n|^2}},
\end{equation}
where $n\in \mathds{C}$ and $0\leq |n|^2\leq 1$. 
We consider  Eq. (\ref{threequbit3}) as the initial state as it covers both GHZ and NMM states. Specifically, for $p=1/2$, the state is nothing but a LU equivalent to conventional GHZ-state. Inverting Eq. (\ref{basis}), we get,
\begin{equation}
\ket{0}=\frac{\ket{+_n}-n\ket{-_n}}{\sqrt{1+|n|^2}},\hspace{2mm} \ket{1}=\frac{n^*\ket{+_n}+\ket{-_n}}{\sqrt{1+|n|^2}}.
\end{equation}
Putting these in Eq. (\ref{threequbit3}) we get,
\begin{align}
&&\ket{NMM}=\ket{+_n}[N\sqrt{p}\ket{\phi^+
}+Nn^*\sqrt{1-p}\ket{\phi^-}]\nonumber\\
&& + \ket{-_n}[-Nn\sqrt{p}\ket{\phi^+
}+N\sqrt{1-p}\ket{\phi^-}],
\end{align}
where $N=1/{\sqrt{1+|n|^2}}$. When Charlie makes a measurement, the collapsed state between Alice and Bob can be one of the following two states,

\begin{align}
&&\ket{\psi_+}_{AB}=
\frac{1}{\sqrt{p_+}}\big[N\sqrt{p}\ket{\phi^+
}+Nn^*\sqrt{1-p}\ket{\phi^-}\big],\label{1collapsed states}\\
&&\ket{\psi_-}_{AB}=
\frac{1}{\sqrt{p_-}}\big[N\sqrt{1-p}\ket{\phi^-}-Nn\sqrt{p}\ket{\phi^+
}
\big],\label{2collapsed states}
\end{align}
with probabilities $p_+=N^2p+N^2|n|^2(1-p)$ and  $p_-=N^2 p|n|^2 +N^2(1-p)$ corresponding to measurement outcomes $\ket{+_n}$ and $\ket{-_n}$ respectively. It is clear from the structure of the collapsed state that this state can be separable, partially entangled or maximally entangled. As an example,
when, $\sqrt{p}=n^*\sqrt{1-p}$, the collapsed state $\ket{\psi_+}$ is separable. When $\sqrt{p}\neq n^*\sqrt{1-p}$, state is partially entangled, and when $n=0$, the collapsed state is maximally entangled.

We rewrite the states in  Eq. (\ref{1collapsed states}) and Eq. (\ref{2collapsed states}) as
\begin{eqnarray}
\ket{\psi_+}_{AB}=N_1(\ket{00}+n_1\ket{11})\label{pQKD1}\\
\ket{\psi_-}_{AB}=N_2(\ket{00}+n_2\ket{11})\label{pQKD2},
\end{eqnarray}
where, we define, 
\begin{align}
\nonumber
\frac{N_1}{N} =\frac{\sqrt{p}+n^\ast\sqrt{1-p})}{\sqrt{2p_+}}, n_1 = \frac{\sqrt p-n^\ast\sqrt{1-p}}{\sqrt p+n^\ast\sqrt{1-p}},\\
\frac{N_2}{N} = \frac{\sqrt{1-p}-n\sqrt p)}{\sqrt{2p_-}}, n_2 = \frac{-\sqrt{1-p}-n\sqrt p}{\sqrt{1-p}-n\sqrt {p}}.
\end{align}

With these partially entangled states Eqs. (\ref{pQKD1},\ref{pQKD2}), Alice and Bob initiate an entanglement based QKD protocol like in Ref. \cite{Ekert}. Let us recall what happens in the original protocol. The original protocol \cite{Ekert,gisinrmp} involves a maximally entangled state, namely a Bell state. In the absence of an eavesdropper, {\it e.g.} Eve, the protocol is reminiscent of the BB84 protocol. Two parties hold one qubit each and agree on two sets of basis states in which they measure their own qubits randomly. After the measurement step, they announce their choices of the bases. Data is kept, if the bases match, otherwise data is discarded. In half of the cases,  Alice and Bob get perfectly correlated results and they can construct a secure key. The secure key rate is $1/2$ in this scenario. 
But often in a practical situation, the perfect correlation is not obtained, which indicates noise in the entanglement channel or imperfect measurement or possibly Eve's intervention. Alice and Bob use a part of the matched data to determine the Quantum Bit Error Rate (QBER) and the remaining part is used to build secure key after error correction and privacy amplification. To know Eve's presence, in the Ekert's original protocol,  Alice and Bob use one extra set of basis states to measure that helps in testing the violation of Bell-CHSH inequality. If it is maximally violated then there is no eavesdropper's attack. Non-maximal violation would indicate a non-zero QBER. In this protocol, with one extra measurement each by Alice and Bob, the key rate changes from $1/2$ to $2/9$. As the resource state in our case is a non-maximally entangled state, we always get a non-maximal violation. So, we have to rearrange the protocol a bit. In the original protocol Alice and Bob choose three measurement settings each, say, $(A_1,A_2, A_3)$ and  $(B_1,B_2, B_3)$ respectively, giving rise to nine combinations of measurement operators. 
 Alice's the settings are : $A_1=\sigma_x$, $A_2=1/\sqrt{2}(\sigma_x+\sigma_y)$ and $A_3=\sigma_y$. Whereas, Bob's measurement settings are : $B_1=1/\sqrt{2}(\sigma_x+\sigma_y)$, $B_2=\sigma_y$ and $B_3=1/\sqrt{2}(-\sigma_x+\sigma_y)$. 
Two combinations used to make the secret key are $(A_2,B_1)$ and $(A_3,B_2)$, because of perfect correlations in these two settings. Four combinations e.g $(A_1,B_1)$, $(A_1,B_3)$, $(A_3,B_1)$ and $(A_3,B_3)$ are required to test the Bell violation. These particular measurement settings give the maximal violation for a Bell state. Remaining combinations of the measurements are discarded. \\\\
\textbf{\textit{Security of the key :}} Let us now consider our case of NMM-state, and see how Bell-CHSH inequality can be used for the security of key generation. 
In our case, we have to choose the measurement settings according to the structure of the resource state. It is well known \cite{horo_bell,scarani2} that given a non-maximally entangled two-qubit pure state, we can always specify the measurement settings for which Bell-CHSH \cite{chsh} inequality is optimally violated. 
The steps in security analysis are following:
First Charlie announces the choice of $n$, which is taken to be a real number, and also the outcome of his measurement.
 Knowing $n$, Alice and Bob know the collapsed state between them, which is either $N_1\ket{00}+N_1n_1\ket{11}$ or
  $N_2\ket{00}+N_2n_2\ket{11}$. Suppose, the collapsed state is $N\ket{00}+Nn\ket{11}$. Then, Alice and Bob choose the following three measurement settings each,  giving rise to nine combinations.
\begin{align}
\nonumber
& A_1=\sigma_z, \hspace{2mm} A_2=\cos\theta\sigma_z+\sin\theta\sigma_x,\hspace{2mm} A_3=\sigma_x\\
\nonumber
& B_1=\cos\theta\sigma_z+\sin\theta\sigma_x,\hspace{1mm} B_2=\sigma_x,\hspace{1mm} B_3=\cos\theta\sigma_z-\sin\theta\sigma_x,
\end{align}
where, $\cos\theta= 1/\sqrt{1+4n^2N^4}$. Like before, the combination $(A_2,B_1)$ and $(A_3,B_2)$ can be used to generate the secret key and the combination $(A_1,B_1)$, $(A_1,B_3)$, $(A_3,B_1)$ and $(A_3,B_3)$ can give the optimal Bell violation, which is $2\sqrt{1+4n^2N^4}$. Remaining results are thrown out. So, if the Bell violation is less than the optimal value $2\sqrt{1+4n^2N^4}$,  Eve's presence is detected.\\\\
\textit{\textbf{QBER and key rate :}} Let us now find optimal key rate without worrying about security. Then,
Alice and Bob would make measurement in only two independent bases.
In the case of a partially entangled state, we have nonzero QBER because we cannot have perfect correlations in two different measurement bases. As stated, in this scenario, Alice and Bob measure with two different choices of basis states. We assume that they have agreed to measure with the projectors $P_{0/1}$ and $P_{\pm}$  where 
\begin{eqnarray}
&&P_{0/1} = M_{0/1}^\dagger M_{0/1},\:\: P_{\pm}=M_{\pm}^\dagger M_{\pm}
\end{eqnarray}
and $M_0=|0\rangle\langle0|$, $M_1=|1\rangle\langle1|$, $M_{\pm}=|\pm_n\rangle\langle\pm_n|$. Here, QBER is defined as the probability that they would obtain different outcomes even if the measurement basis states are same, {\it i.e.}, 
\begin{eqnarray}
QBER&=&\rm{Tr}(P_0 \otimes P_1 \rho)+\rm{Tr}(P_1 \otimes P_0 \rho)+\nonumber\\&
&\rm{Tr}(P_+\otimes P_- \rho)+\rm{Tr}(P_-\otimes P_+ \rho),
\label{qberfirst}
\end{eqnarray}
where $\rho=|\psi\rangle\langle\psi|$. Now plugging $\ket{\psi_+}_{AB}$ into the above equation we find the expression for QBER:
\begin{equation}\label{qber_1}
QBER=\frac{n^2(1-p)}{2\big(n^2(1-p)+p\big)}.
\end{equation}
We have plotted it with Charlie's measurement parameter $n$.
\begin{figure}[h]
\centering
\includegraphics[scale=0.6]{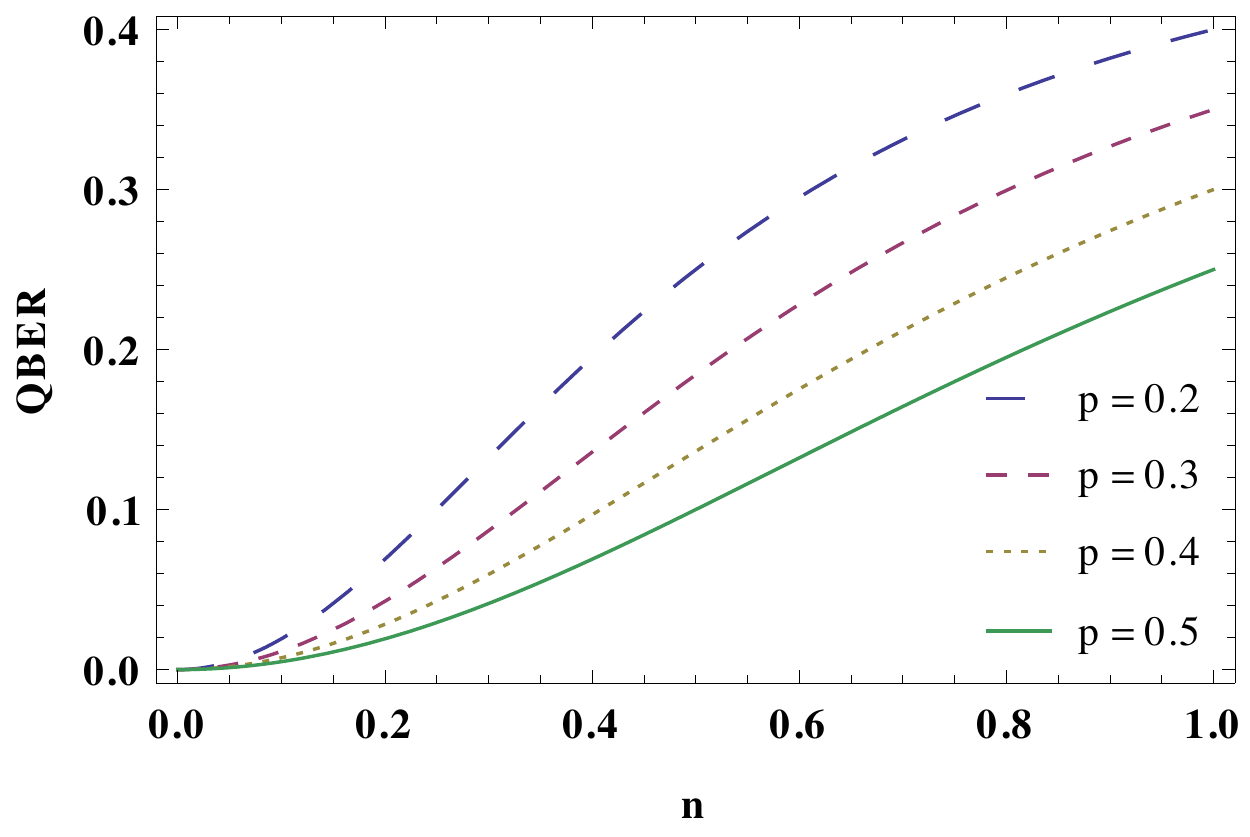}
\caption{QBER with control power of Charlie. Lower one corresponds to GHZ-state for which QBER is least of all.}
\label{fig_qber_n}
\end{figure} 
It shows that error rate vanishes if measurement basis states correspond
 to $n=0$, in which case the collapsed states are Bell states, leading to perfectly correlated measurement outcomes. In more general scenario, {\it i.e.,} when the state is not a Bell state, there exists error rate even in the absence of any eavesdropper. Interestingly,
  Charlie can control QBER by choosing appropriate measurement basis states.  
Now, with information about QBER, Alice and Bob can employ some error correcting protocols to distill a secure key. Having calculated the QBER, next thing is to determine the key rate. In the absence of Eve and with no QBER, the sifted key rate is $1/2$. But the final key rate depends on the practical implementation of the protocol and hence on QBER. So, a more reasonable quantity to use is the relative key rate which is determined by the difference between Bob's and eavesdropper Eve's information about sifted key \cite{gisinrmp,aho,scarani1}. The relative key rate is defined as,
$r=R_{fin}/R_{sif}=I(A,B)-I(A,E)$ (for symmetric individual attacks), where $R_{fin}$, $R_{sif}$ are the final and sifted key rate respectively and $I(A,B)$ is the mutual information between Alice and Bob, and similarly, $I(A,E)$ is the mutual information between Alice and Eve. In our case, we are neglecting the presence of Eve. So, we have \cite{gisinrmp,aho,scarani1}, $r=1+Q\log_2 Q+(1-Q)\log_2 (1-Q)$, where $Q$ is the QBER.
After the conclusion of the protocol, Alice and Bob inform Charlie about the key rate and QBER. If there is a discrepancy between the expected QBER (and key rate) and reported QBER (and key rate), then Charlie knows that there is cheating and that run of secret key making is rejected. Moreover, without Charlie's measurement Alice and Bob cannot establish a key, as their qubits are in maximally mixed states. So, the whole key making process is initiated and supervised by a controller/supervisor, Charlie, and thus making the protocol secure and trustworthy.
\\\\
{\em \textbf{Remarks}}.-- In the above, we see that if, the reduced density matrices of Alice and Bob have entropy one from the beginning, then by making a measurement in a right basis, Charlie can reduce the state between Alice and Bob to a maximally entangled state. So, the question arises that, if the qubits held by Alice and Bob do not have entropy one ({\it i.e.}, maximally mixed) but less than one from the beginning, can a measurement by Charlie make them maximally mixed? In the following, we show that this is not possible. To show this, we start with the state of the form of Eq. (\ref{threequbit3}). But this time Charlie makes a measurement on a qubit, which has entropy one. Then we show that it is not possible to increase the entropy of the qubit, which has entropy less than one.  Considering the state in Eq. (\ref{threequbit3}), let us say, Charlie makes a measurement on the second qubit in the general basis as given in Eq. (\ref{basis}). In terms of this basis, we can write the state as,
\begin{align}
\label{notone}
\ket{NMM} = N/\sqrt{2}\Big[\sqrt{p}\ket{00}+n^*\sqrt{p}\ket{01} \nonumber \\
 +\sqrt{1-p}\ket{10}-n^*\sqrt{1-p}\ket{11}\Big]\ket{+}\nonumber  \\
  +N/\sqrt{2}\Big[-n\sqrt{p}\ket{00}+\sqrt{p}\ket{01} \nonumber \\
  -n\sqrt{1-p}\ket{10}-\sqrt{1-p}\ket{11}\Big]\ket{-}.
\end{align}
Corresponding to two outcomes of Charlie's measurement, the collapsed state between Alice and Bob is,
\begin{eqnarray}
\label{notone2}
\nonumber
 \ket{\phi^+} & = & \sqrt{p}\ket{0}\Big[N[\ket{0}+n^*\ket{1}]\Big]\\
& & +\sqrt{1-p}\ket{1}\Big[N[\ket{0}-n^*\ket{1}]],\\
\nonumber
 \ket{\phi^-} &= &\sqrt{p}\ket{0}\Big[N[-n\ket{0}+\ket{1}]\Big]\\
& & +\sqrt{1-p}\ket{1}\Big[N[-n\ket{0}-\ket{1}]].
\end{eqnarray}
It is evident from the expression that, for
the collapsed state to be a Bell state, we must have $n=1$ and $p=1/2$. This eventually makes the 
starting state to be a GHZ state. Maximum entropy we can achieve for the qubit held by Bob is what we had
 from the beginning and that can be attained when $n=1$. This shows that we must have atleast two qubits 
 to have entropy one, on which measurements are not being done. We can do CoQKD with a partially 
 entangled state, but then key rate would not be optimal.
\section{Conference QKD with NMM-state}\label{sec3}
In the previous section, we discussed the controlled QKD scheme, where one party's role was to do the measurement and supervise the establishment of a secret key between the remaining two parties. 
In this section, we discuss the protocol for establishing a secret key among all three parties, also called a conference key, with the NMM-state as a resource state. There are several conference key protocols using multipartite entanglement \cite{cabello,scapra,lo,bruss}. we are adopting the scheme introduced in the Ref. \cite{scapra}. We show that one can generate  conference key using  the NMM-state with some non-zero QBER. We first illustrate how the protocol works for GHZ-state. Before going to describe the protocol, first we note the following equivalences, without invoking any local unitary, 
\begin{align}
\sqrt{1/2}\big[\ket{+_x}\ket{\phi^+}+\ket{-_x}\ket{\phi^-}\big]=
\sqrt{1/2}(\ket{000}+\ket{111})\nonumber\\
=\sqrt{1/2}\big[\ket{+_y}\ket{\Phi^-}+\ket{-_y}\ket{\Phi^+}\big]\nonumber~~~~~~~~~~~~
\end{align}
where, $\ket{+_x}=\sqrt{1/2}(\ket{0}+\ket{1})$ and $\ket{-_x}=\sqrt{1/2}(\ket{0}-\ket{1})$ are the eigenstates of $\sigma_x$, $\ket{+_y}=\sqrt{1/2}(\ket{0}+i\ket{1})$ and $\ket{-_y}=\sqrt{1/2}(\ket{0}-i\ket{1})$ are the eigenstates of $\sigma_y$, $\ket{\Phi^+}= \sqrt{1/2}(\ket{00}+i\ket{11})$ and $\ket{\Phi^-}= \sqrt{1/2}(\ket{00}-i\ket{11})$. 
One can also show that,
\begin{align}
\label{bellchange}
\ket{\phi^+}&=\sqrt{1/2}(\ket{+_x}\ket{+_x}+\ket{-_x}\ket{-_x}\nonumber\\
&=\sqrt{1/2}(\ket{+_y}\ket{-_y}+\ket{-_y}\ket{+_y}).\\
\ket{\Phi^+}&=\sqrt{1/2}(\ket{+_x}\ket{+_y}+\ket{-_x}\ket{-_y})\nonumber\\
&=\sqrt{1/2}(\ket{+_y}\ket{+_x}+\ket{-_y}\ket{-_x}),
\end{align}
and similarly for $\ket{\phi^-}$ and $\ket{\Phi^-}$.
Above equations show that whenever odd number of $\sigma_x$ and even number of $\sigma_y$ are measured, one gets perfect correlations. This is because the GHZ state is the simultaneous eigenstate of the stabilizer group containing eight elements, \cite{lo} which are $\{III, XXX, ZZI, IZZ, ZIZ, -YXY, -YYX, XYY\}$.
This observation is crucial for the protocol of conference QKD, established between Alice, Bob, and Charlie.
To start the protocol, first Alice, Bob and Charlie have one qubit each.
These three qubits are in GHZ-state. All three parties make measurement in either  $\sigma_x$ or $\sigma_y$ basis randomly. For the time being, we are not concerned with Eve's presence. Then all of them  publicly announce their measurement basis choices. They keep the data for which all of them measure $\sigma_x$ or any two of them measure $\sigma_y$. They discard the remaining data. As, their results are now perfectly correlated, they can generate a secret key. Out of eight set of measurements, they  keep four of them to generate the key. So, the key rate is $1/2$.

Now, we consider the NMM-state. We take the starting form of the state to be $\ket{\psi}=\sqrt{p}\ket{+_x}\ket{\phi^+}+\sqrt{1-p}\ket{-_x}\ket{\phi^-}$. As expected, we would not get perfect correlations;  so QBER is not zero. We calculate  QBER for this kind of state. To do that, we write down the correlation and anti-correlation charts for different choices of measurements as shown in Table \ref{corrtable}. The charts show the possible outcomes when different combinations of measurements are done.   
\begin{table}
 \centering
\begin{tabular}{|ccc|}
\hline
{$\sigma_x$}  &   $\sigma_x$  &  $\sigma_x$\\
\hline
$+$ & $+$ & $+$\\
$+$ & $-$ & $-$\\
$-$ & $+$ & $-$\\
$-$ & $-$ & $+$\\
\hline
\end{tabular}
\quad
\begin{tabular}{|ccc|}
\hline
{$\sigma_x$}  &   $\sigma_y$  &  $\sigma_y$\\
\hline
$+$ & $+$ & $-$\\
$+$ & $-$ & $+$\\
$-$ & $+$ & $+$\\
$-$ & $-$ & $-$\\
\hline
\end{tabular}
\quad
\begin{tabular}{|ccc|}
\hline
{$\sigma_y$}  &   $\sigma_x$  &  $\sigma_y$\\
\hline
$+$ & $+$ & $-$\\
$+$ & $-$ & $+$\\
$-$ & $+$ & $+$\\
$-$ & $-$ & $-$\\
\hline
\end{tabular}
\quad
\begin{tabular}{|ccc|}
\hline
{$\sigma_y$}  &   $\sigma_y$  &  $\sigma_x$\\
\hline
$+$ & $+$ & $-$\\
$+$ & $-$ & $+$\\
$-$ & $+$ & $+$\\
$-$ & $-$ & $-$\\
\hline
\end{tabular}
\caption{Correlation and anti-correlation tables.}
\label{corrtable}
\end{table}

Note that the starting state, $\ket{\psi}$ is LU equivalent to the NMM-state. The state has perfect correlations for three $\sigma_x$ measurements. So, QBER is zero for this kind of correlation. Also, when $\sigma_x$ is measured on the first qubit,  it is straightforward to see that,
\begin{eqnarray}
\nonumber
& \sqrt{p}\ket{+_x}\ket{\phi^+}+\sqrt{1-p}\ket{-_x}\ket{\phi^-}=\\
&\sqrt{p/2}\ket{+_x}(\ket{+_y}\ket{-_y}+\ket{-_y}\ket{+_y})+\\
& \sqrt{(1-p)/2}\ket{-_x}(\ket{+_y}\ket{+_y}+\ket{-_y}\ket{-_y}).
\end{eqnarray}

So, for the the measurements $XXX$ and $XYY$, we have perfect correlations and hence zero QBER. But, the other two measurements do not give perfect correlations. From the chart of correlation and anti-correlation it is evident that remaining two cases {\it i.e.},  $YXY$ and $YYX$  give same QBER. Let us compute it for the first one. 
For the table of $YYX$, the QBER is given by,
\begin{align}
\nonumber
&Tr[(P^y_+\otimes P^y_+\otimes P^x_+)\rho]+Tr[(P^y_+\otimes P^y_-\otimes P^x_-)\rho]\\
\nonumber
&+Tr[(P^y_-\otimes P^y_+\otimes P^x_-)\rho]+Tr[(P^y_-\otimes P^y_-\otimes P^x_+)\rho],
\end{align}
which comes out to be equal to $1/2(\sqrt{1-p}-\sqrt{p})^2$ and same for the other table. The expression of QBER is obtained from the correlation and anti-correlations charts by writing the probability of the combinations of outcomes which are opposite to that written in the charts.\\
we plot the QBER with $p$ and see that as expected, it is zero for GHZ-state.
\begin{figure}[h]
\centering
\includegraphics[scale=0.6]{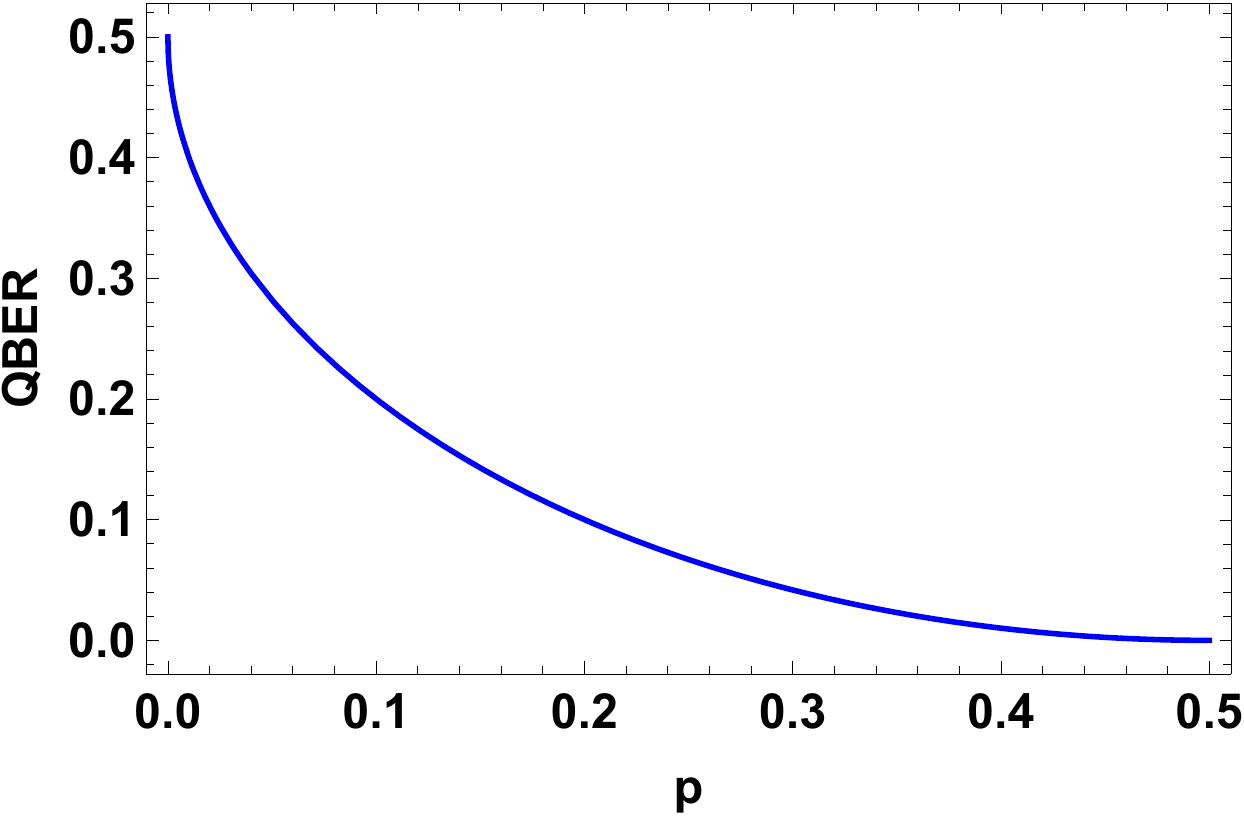}
\caption{Variation of QBER with $p$}
\label{conference}
\end{figure}
We see that the NMM-state is useful for conference QKD scheme, but with some non-zero QBER. The security of the secret key in the presence of Eve can be analyzed by estimating her maximal information gain. If Eve has one qubit as a resource, then with the help of that, she can affect both the channels $A-B$ or $A-C$ \cite{scapra}, with the assumption of individual attack,  {\it i.e.} at a time she can affect one channel only. We also assume that all the three parties between whom the secret key is being established are trusted. It is only an outsider like Eve who wants to jeopardize the protocol. We note this line of attack can take place if Alice has
all the three qubits in the beginning. She put these three qubits in GHZ state and then sends one qubit each to Bob and Charlie. Estimation of $I(A;E)$ w.r.t $I(A;BC)$ gives the maximal knowledge that Eve can acquire about the protocol and if $I(A;BC)>I(A;E)$ then only the all three partners can go on to make the secret key. This is guaranteed from the violation of MABK inequalities. With Eve's qubit and her maximal power of affecting the protocol, we can construct the total four-qubit state $\ket{\Psi_{ABCE}}$, with $\rho_{ABC}$ and $\rho_{AE}$, being the trusted and untrusted part respectively. In \cite{scapra} it was shown that $I(A;BC)$ is greater than $I(A;E)$, only when $\rho_{ABC}$ violates the MABK \cite{mermin,ardehali} inequality but not $\rho_{AE}$. In the same way, starting with the state $\sqrt{p}\ket{+_x}\ket{\phi^+}+\sqrt{1-p}\ket{-_x}\ket{\phi^-}$, one can first analyze the maximal attack possible by Eve and determine the state $\ket{\Psi_{ABCE}}$. Then one can go on to check condition for $I(A;BC)>I(A;E)$ in terms of Bell inequality violation. There is a zoo of Bell inequalities \cite{brunnerrmp} in multipartite scenario. One has to choose the optimum Bell inequality for this scenario, which can be used most effectively. In this sense, the inequalities \cite{arpan,arpan1} previously introduced by some of the present authors are useful, as these inequalities are violated by all generalized GHZ-states, the property which is not shown by any correlation Bell inequalities with two dichotomic measurement choices per party. In the following, we show the protocol in which using the inequalities introduced in \cite{arpan}, we can detect the presence of Eve. We take any inequality out of the set of six inequalities constructed in \cite{arpan}, and see the violation by the NMM-state. We take the following inequality, 
\begin{equation}
I=A_1(B_1+B_2)+A_2(B_1-B_2)C_1\leq 2.
\label{ourin}
\end{equation}
In the following we show that every NMM-state violates the inequality for the whole parameter range. To show this, we choose the following measurement settings,
\begin{align}
& A_1=\sigma_z, \hspace{2mm}A_2=\sigma_x\\
\nonumber
& B_1=\cos\theta\sigma_x+\sin\theta\sigma_z, \hspace{2mm} B_2=-\cos\theta\sigma_x+\sin\theta\sigma_z\\
\nonumber
&C_1=\sigma_x
\end{align}
These measurement settings are similar to the one used in Ref. \cite{arpan}. For these measurement settings, the expectation value of the state $\ket{\psi}=\sqrt{p}\ket{+_x}\ket{\phi^+}+\sqrt{1-p}\ket{-_x}\ket{\phi^-}$ is $\langle I\rangle=4\sqrt{p(1-p)}\cos\theta +2\sin\theta$. Employing the inequality $\alpha \sin\theta +\beta \cos\theta\leq \sqrt{\alpha^2+\beta^2}$, it is evident that, $\langle I \rangle\leq 2\sqrt{1+4p(1-p)}$. This shows that the inequality always gets violated by the NMM-state and when $p=1/2$ ( {\it i.e.} GHZ-state), the violation is $2\sqrt{2}$. Therefore, for $\cos\theta=1/\sqrt{1+4p(1-p)}$, the measurement settings we chose is also the optimal measurement settings for the NMM-state. Next, we describe the protocol to detect the presence of Eve. For this, in each round of the protocol, Alice chooses from three measurement settings, Bob chooses from four measurement settings and Charlie chooses from three measurement settings as following,
\begin{align}
& A_1=\sigma_x, \hspace{2mm} A_2=\sigma_y,\hspace{2mm} A_3=\sigma_z\\
\nonumber
&  B_1=\sigma_x,\hspace{1mm}  B_2=\sigma_y,\\
\nonumber
& B_3=\cos\theta\sigma_x+\sin\theta\sigma_z,\hspace{1mm} B_4=-\cos\theta\sigma_x+\sin\theta\sigma_z.\\
\nonumber
& C_1=\sigma_x, \hspace{2mm} C_2=\sigma_y,\hspace{2mm} C_3=\mathbb{I},
\end{align}
where, $\mathbb{I}$ is an Identity operator, and $\cos\theta=\frac{1}{\sqrt{1+4p(1-p)}}$. So, there are total $36$ combinations, out of which four combinations {\it e.g.} $(A_1, B_1, C_1)$, $(A_1, B_2, C_2)$, $(A_2,B_1,C_2)$ and $(A_2, B_2, C_1)$ are used to make the key as described before. This gives the key rate of $1/9$. Four combinations,  {\it e.g.} $(A_3, B_3, C_3)$, $(A_3, B_4, C_3)$, $(A_1,B_3,C_1)$ and $(A_1, B_4, C_1)$ are used to  check the optimal violation for the inequality in Eq. (\ref{ourin}), which is $2\sqrt{1+4p(1-p)}$. So, if the expectation value for the inequality is less than $2\sqrt{1+4p(1-p)}$, we can surely say that there is Eve's intervention. Remaining $28$ combinations we throw away for the completion of the protocol.
\section{Resource states structure for multipartite case ($N\ge 4)$}\label{sec4}
In this section, we explore the resource state structure of four-qubit states that are suitable for maximal CoQKD, {\it i.e.} CoQKD with perfect key and optimal key rate,
 for some particular measurements performed by the controllers. In this multipartite protocol, all the controllers are independent and they announce their choices of bases publicly after making the measurements.
 In the scenario of four parties, there may arise two situations for CoQKD. In the first case, there are four parties and each party has one qubit. Second case is when there are three parties and one party (other than the sender and the receiver) has two qubits.
We start with the first case.
\subsection{Case I: Each party has one qubit}
Secret key can be established between Alice and Bob, if after the measurements by Charlie and Dennis, they share a Bell state or its LU equivalent state. \\\\
\textit{\textbf{Proposition-2}} : \textit{For maximal {Controlled} QKD, after the measurement by one party say Dennis, the collapsed state between Alice, Bob and Charlie is LU equivalent to \\
$\sqrt{1/2}\big[\ket{0}\ket{\phi^+}+\ket{1}(\mathds{1}\otimes U)\ket{\phi^-}\big]$. If $(\mathds{1}\otimes U)\ket{\phi^-}$ is orthonormal to $\ket{\phi^+}$, then it is LU equivalent to GHZ-state.}\\\\
\textit{Proof} : The proof follows from the Proposition-1. There we showed that most general three-qubit resource state structure, for maximal CoQKD, is such that the state
is either LU equivalent to $GHZ$ or LU equivalent to\\
$\sqrt{1/2}\big[\ket{0}\ket{\phi^-}+\ket{1}(\mathds{1}\otimes U)\ket{\phi^+}\big]$, such that $(\mathds{1}\otimes U)\ket{\phi^-}$ is not orthonormal to $\ket{\phi^+}$. Similarly, for four qubits, after the measurement by one party, the state must collapse to one of these three-qubit states,

\begin{align}
\ket{\Phi_1} = \frac{1}{\sqrt{2}}(\ket{0}_4\otimes \ket{g_0}_{321}+ \ket{1}_4\otimes \ket{g_1}_{321}).\label{resource1}\\
\ket{\Phi_2} =\frac{1}{\sqrt{2}}(\ket{0}_4\otimes \ket{g_0}_{321}+ \frac{1}{\sqrt{2}}\ket{1}_4\otimes \ket{\psi}_{321}).\label{resource2}\\
\ket{\Phi_3} = \frac{1}{\sqrt{2}}(\ket{0}_4\otimes \ket{\psi}_{321}+ \frac{1}{\sqrt{2}}\ket{1}_4\otimes \ket{\psi'}_{321})\label{resource3}.
\end{align}
%
where, $\ket{g_0}$ is conventional GHZ-state and 
$\ket{g_1}$ is LU equivalent  $\ket{g_0}$, $\ket{\psi}$ and $\ket{\psi'}$ are the states as mentioned in Eq. (\ref{threequbit3}) with different coefficients $p$. The subscripts denote the order of the qubits which we follow throughout our discussion. 
Notice that, if $\ket{g_0}$ is orthogonal to $\ket{g_1}$, then all the single qubit reduced density matrices of the resource state $\ket{\Phi_1}$ have entropy one, otherwise, 
the reduced density matrix of the first qubit has entropy less than one, whereas all other single qubit reduced density matrices have entropy one. For the state $\ket{\Phi_2}$, the single qubit reduced density matrices for the last two qubits are maximally mixed, whereas the other qubits have reduced density matrices with entropy less than one. And similarly the state $\ket{\Phi_3}$ has also the similar entropy configuration as $\ket{\Phi_2}$.
Therefore, given the entropy structures, we cannot distinguish between $\ket{\Phi_2}$ and $\ket{\Phi_3}$. To distinguish them, we need the collapsed states after the measurement by Dennis. $\square$
 
To illustrate the above proposition, we give examples of states which 
satisfy the above resource structure. Cluster states \cite{cluster} which belong to the first category of states with all the single qubit reduced density matrices with entropy one are  suitable resource states for CoQKD. A nontrivial example is the following state, 
\begin{align}
\ket{R_1} = \frac{1}{2\sqrt2}\Big[|0010\rangle+|0100\rangle+|0001\rangle-|0111\rangle+\nonumber\\
|1000\rangle+|1100\rangle+|1011\rangle
 -|1111\rangle\Big],\nonumber
\end{align}
This state has similar entropic structure like the first structure, where $\ket{g_0}$ and $\ket{g_1}$ are not orthogonal as  $S(\rho_4)\approx.81$ and $S(\rho_i)=1$ for $i=1,2,3$. This state is also suitable for {controlled} QKD.
For four-qubits, we see that two maximally-mixed single qubit reduced density matrices are the minimum requirement. Therefore, for four-qubits, this criteria is also a necessary and sufficient for successful {controlled} QKD. Other possible structures are sufficient conditions. So, we can generalize it for any $N$-qubit entangled state with each one holding a single subsystem.\\\\
\textit{\textbf{Proposition}} : \textit{For maximal controlled QKD, using a $N$-qubit entangled state, with each party holding one qubit, the necessary and sufficient condition for the resource state is that it has at least two maximally-mixed single qubits}.

\subsection{Case II: One party has more than one qubit}

Let us now consider the second situation, where one party, {\it e.g.} Charlie, has two qubits in his possession. 
We know that for CoQKD the state must be either NMM (necessary and sufficient) or GHZ (sufficient) along the cut Charlie-(Alice and Bob). Now, Charlie holds two qubits, say first two qubits, and makes measurement in the orthonormal computational basis, $\{\ket{00},\ket{01},\ket{10},\ket{11}\}$. We require that after the joint measurement, the collapsed state between Alice and Bob is  LU equivalent state to a Bell state. Therefore, the most general structure of the resource state is,
\begin{eqnarray}
\nonumber
&&\ket{\Psi}=\frac{1}{2}(\ket{00}\otimes \ket{\phi_1}+
\ket{01}\otimes \ket{\phi_2}\\
&&+\ket{10}\otimes \ket{\phi_3}+\ket{11}\otimes \ket{\phi_4}),\label{four2}
\end{eqnarray}
where, $\ket{\phi_1}$, $\ket{\phi_2}$, $\ket{\phi_3}$ and $\ket{\phi_4}$ are LU equivalent to Bell states, though not necessarily orthonormal to each other. All states of the form of Eq. (\ref{resource1}), Eq. (\ref{resource2}) and Eq. (\ref{resource3}) can be recast as Eq. (\ref{four2}) and vice versa. 
It can be shown very easily, by observing that $\sqrt{1/2}(\ket{0}\otimes\ket{\phi_1}+\ket{1}\otimes\ket{\phi_2})$ and $\sqrt{1/2}(\ket{0}\otimes\ket{\phi_3}+\ket{1}\otimes\ket{\phi_4})$ are the NMM states. So, $\ket{\Psi}$ has the similar structure like $\ket{\Phi_1}$, $\ket{\Phi_2}$ and $\ket{\Phi_3}$.
Therefore, the states which are useful for CoQKD as described in the case I, can also be used as a resource in the present case II. 
%
\subsection{QBER for the state between Alice and Bob}
As the structure of the resource states are similar for two scenarios, we can start with any of the above written forms of the resource states. We choose the form of the state to be Eq. (\ref{four2}).
First we start with the scenario, when each party has one qubit with them. In this case there are two controllers/supervisors {\it e.g.} Charlie and Dennis.
As before in the three-qubit scenario, collapsed state between Alice and Bob depends upon the measurement basis chosen by the controllers.

\begin{figure}[h]
\centering
\includegraphics[scale=0.62]{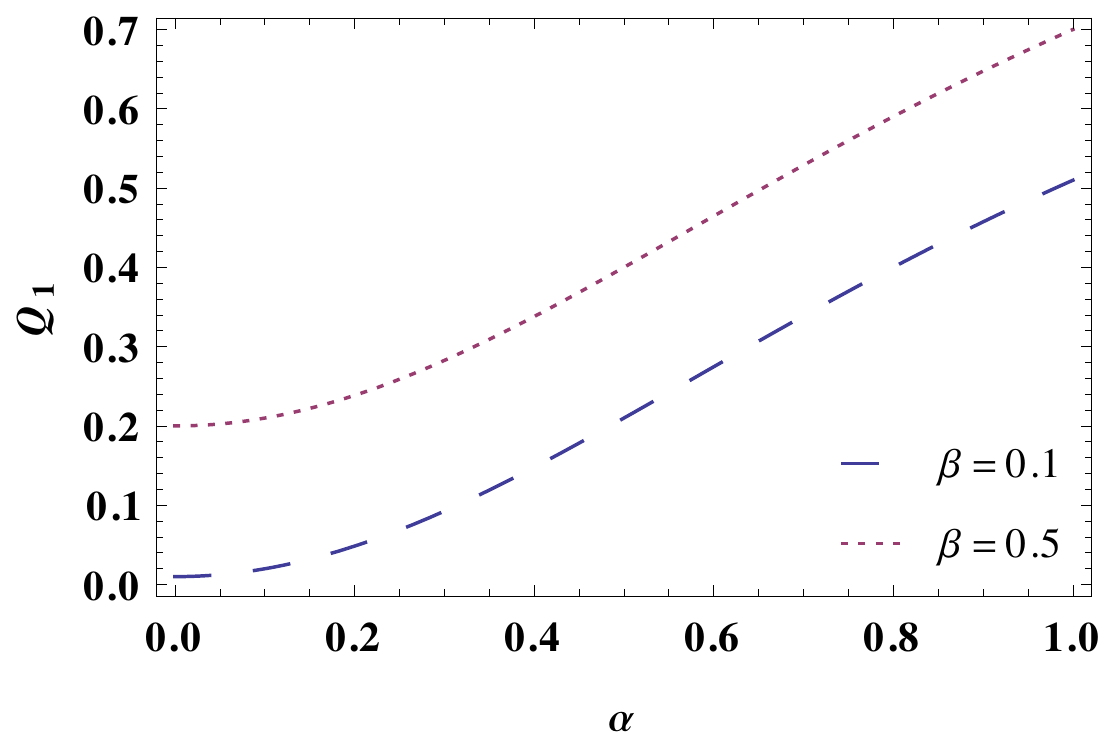}
\caption{Variation of QBER with $\alpha$ for different $\beta$'s.}
\label{qfour}
\end{figure}

Without the loss of generality, we consider that Dennis  and Charlie choose to measure in the basis given by Eq. (\ref{basis}). 
After the measurement by Dennis, the collapsed state between Charlie, Alice and Bob corresponding to the outcome $\ket{+_\beta}$, is given by,
\begin{equation}\label{collapsed_danish}
\ket{\Psi^+_{CAB}}=\frac{\ket{0}(\ket{\phi_1}+\beta\ket{\phi_3})+\ket{1}(\ket{\phi_2}+\beta\ket{\phi_4})}{\sqrt{2(1+|\beta|^2})}
\end{equation}
where $\beta$ determines the measurement basis chosen by Denis.
Thus the collapsed state is a partially entangled three-qubit state. So there is non-vanishing QBER in the protocol. We can find it by using the same procedure as given in Eq. (\ref{qberfirst}). It is to be noted that there is a collapsed state corresponding to the measurement outcome $\ket{-_\beta}$ but the analysis is similar to the present case. Now, Charlie would measure in the general basis as before resulting a collapsed state between Alice and Bob. Corresponding to the outcome $\ket{+_\alpha}$, we find that the state is given by,
\begin{equation}\label{collapsed_charlie}
\ket{\Psi^+_{AB}}=\frac{\ket{\phi_1}+
\alpha\ket{\phi_2}+\beta\ket{\phi_3}+ \alpha\beta\ket{1}\ket{\phi_4}}{\sqrt{(1+|\alpha|^2)(1+|\beta|^2)}},\\
\end{equation} 
where $\alpha$ is the measurement parameter of Charlie and Charlie obtains the outcome $\ket{+_\alpha}$. Now, the collapsed state between Alice and Bob involves two parameters arising from the measurement basis of Charlie and Dennis. We take these two parameters to be real. As the state is not maximally entangled there is  non-vanishing QBER even in the absence of any eavesdropper. When $\ket{\phi_1}$, $\ket{\phi_2}$, $\ket{\phi_3}$ and $\ket{\phi_4}$ are four Bell states, we find the QBER of the protocol employing the expression of Eq. (\ref{qberfirst}) as given by, 
\begin{equation}
Q_1=\frac{\beta^2}{1+\beta^2}+\frac{\alpha^2}{1+\alpha^2}
\end{equation}
There are two parameters in QBER which are controlled by Dennis and Charlie. The behavior of QBER with respect to the measurement parameters are displayed in Fig. \ref{qfour}. We have plotted the QBER with the parameter $\alpha$, for different values of $\beta$'s.
From the expression of QBER, it is evident that plot with $\beta$ for different $\alpha$'s is similar. 
We can see from Eq. (\ref{collapsed_danish}) and Eq. (\ref{collapsed_charlie}) that if Dennis and Charlie measure in the computational basis the collapsed states are three-qubit GHZ-state and Bell state respectively. Therefore, we find vanishing QBER which can be seen in the above plot.

Next, we consider the second scenario, where there is one controller {\it e.g.} Charlie, who holds two qubits. Charlie now can use an entangled basis for the measurement. We show that for the measurement in entangled basis, the collapsed state between Alice and Bob is not a Bell state. 
For the resource state in Eq. (\ref{four2}), Charlie performs a joint measurement using Generalized Bell Basis (GBS), 

\begin{eqnarray}\label{GBS}
\ket{\chi_m^+}=\frac{\ket{00}+m\ket{11}}{\sqrt{1+|m|^2}},\:\:
\ket{\chi_m^-}=\frac{m^\ast\ket{00}-\ket{11}}{\sqrt{1+|m|^2}},\nonumber\\
\ket{\zeta_m^+}=\frac{\ket{01}+m\ket{10}}{\sqrt{1+|m|^2}},\:\:
\ket{\zeta_m^-}=\frac{m^\ast\ket{01}-\ket{10}}{\sqrt{1+|m|^2}},
\end{eqnarray}

\begin{figure}[h]
\centering
\includegraphics[scale=0.55]{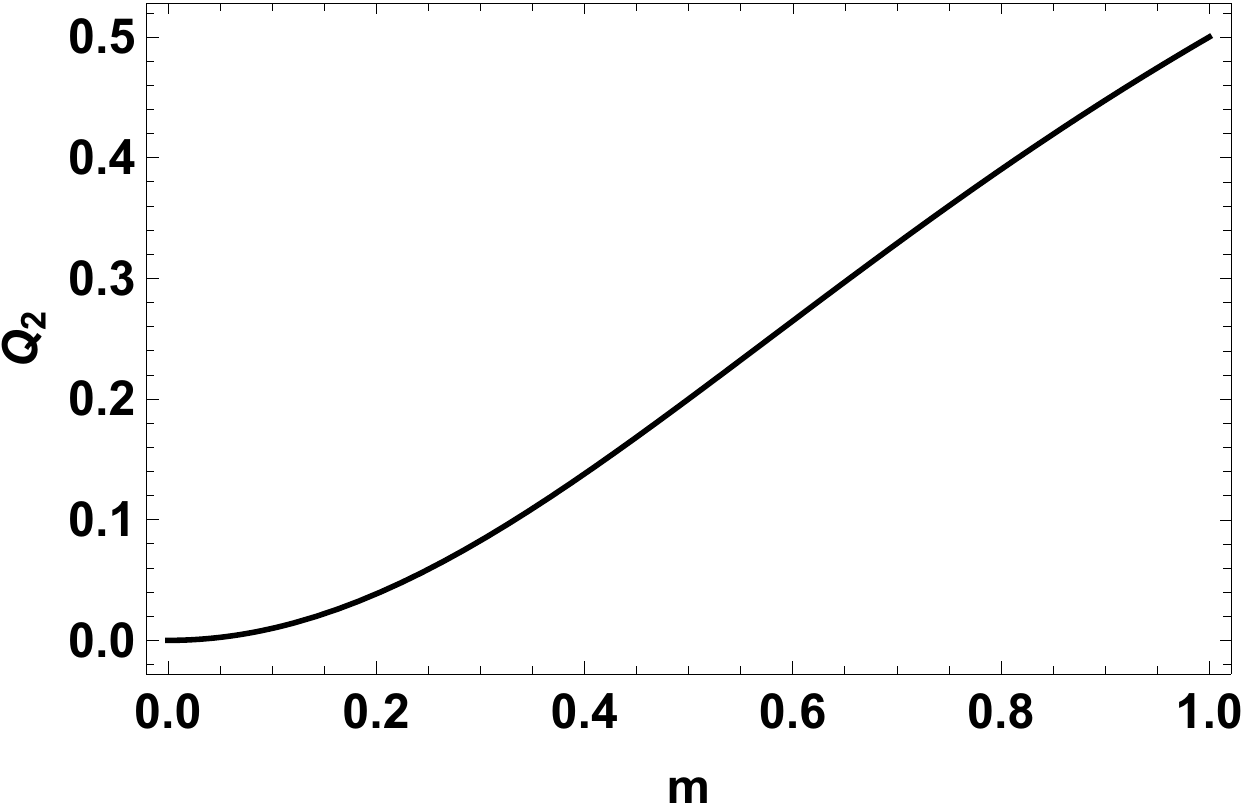}
\caption{Variation of QBER with the parameter $m$.}
\label{fig_case2}
\end{figure}

Then, the collapsed state between Alice and Bob is
\begin{equation}
\ket{\Psi^+_{AB}}=\frac{\ket{\phi_1}+m\ket{\phi_4}}{\sqrt{1+m^2}},
\end{equation}
 corresponding to the outcome $\ket{\chi_m^+}$, which occurs with probability $1/4$. The collapsed state is partially entangled state and as before we find QBER of the protocol,
\begin{equation}
Q_2= \frac{m^2}{1+m^2},
\end{equation}
where, we have considered $m$ as real. If Charlie measures in computational basis the collapsed state is a Bell state which yields a vanishing bit error rate. We plot the QBER in this case with the parameter $m$ in Fig. \ref{fig_case2}.

\section{Cooperative teleportation}\label{sec5}
In this section, we discuss the usefulness  of the resource state (\ref{threequbit3}) in cooperative teleportation scheme \cite{bennet1,pankaj1,generalisedw1,cotele1}. The protocol can be carried out in the same way as CoQKD: Charlie makes a measurement on his subsystem and classically communicates his measurement outcome to Alice who has the unknown qubit to teleport to Bob. She then makes a joint measurement on the composite system conditioned on Charlie's measurement outcome and informs her measurement outcome to Bob who finds a suitable unitary transformation to retrieve the original state.

In tripartite scenario, the state recast as (\ref{threequbit3}) is suitable for the protocol. In this case, depending on Charlie's choice of measurement basis $\it{i.e.}$ on $n$, fidelity $\mathcal{F}$ of teleportation would be determined and for $p=\frac{1}{2}$, Charlie  has full control over the protocol \cite{s_ghose}. As the collapsed state between Alice and Bob is partially entangled in a more general situation, naturally teleportation fidelity is no longer unity. It would be interesting to find the behavior of average fidelity with the Charlie's measurement outcome which is characterized by $n$. We define average fidelity as,
\begin{equation}
\mathcal{F}_{\mbox{av}}=p_+\mathcal{F}_++p_-\mathcal{F}_-,
\end{equation} 
where, $\mathcal{F}_{\pm}$ corresponds to the fidelity of $\ket{\psi_{\pm}}_{AB}$ respectively. We calculate fidelity using the formula given by Horodecki {\it et al.} \cite{horotele}, $\mathcal{F}\leq \frac{1}{2}(1+\frac{1}{3}\mbox{Tr}\sqrt{T^\dagger T})$, where the elements of the matrix $T$ is defined as $t_{\alpha\beta}=\mbox{Tr}\big[(\sigma_\alpha\otimes
\sigma_\beta)\rho\big]$ for a state $\rho$. We consider the optimal fidelity $\mathcal{F}=\frac{1}{2}(1+\frac{1}{3}\mbox{Tr}\sqrt{T^\dagger T})$.

\begin{figure}[h]
\centering
\includegraphics[scale=0.6]{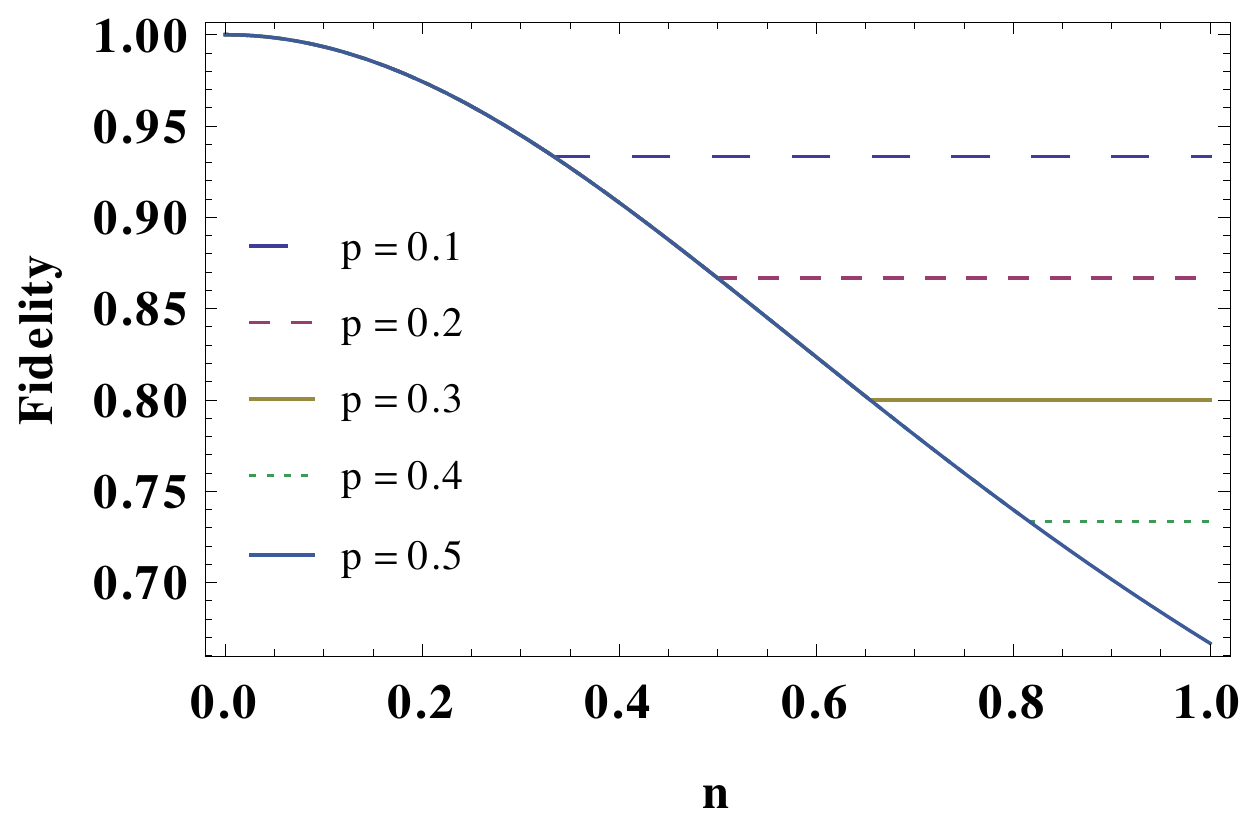}
\caption{Variation of average fidelity with the parameter $n$ }
\label{fig_fidelity}
\end{figure}

\begin{figure}[h]
\centering
\includegraphics[scale=0.6]{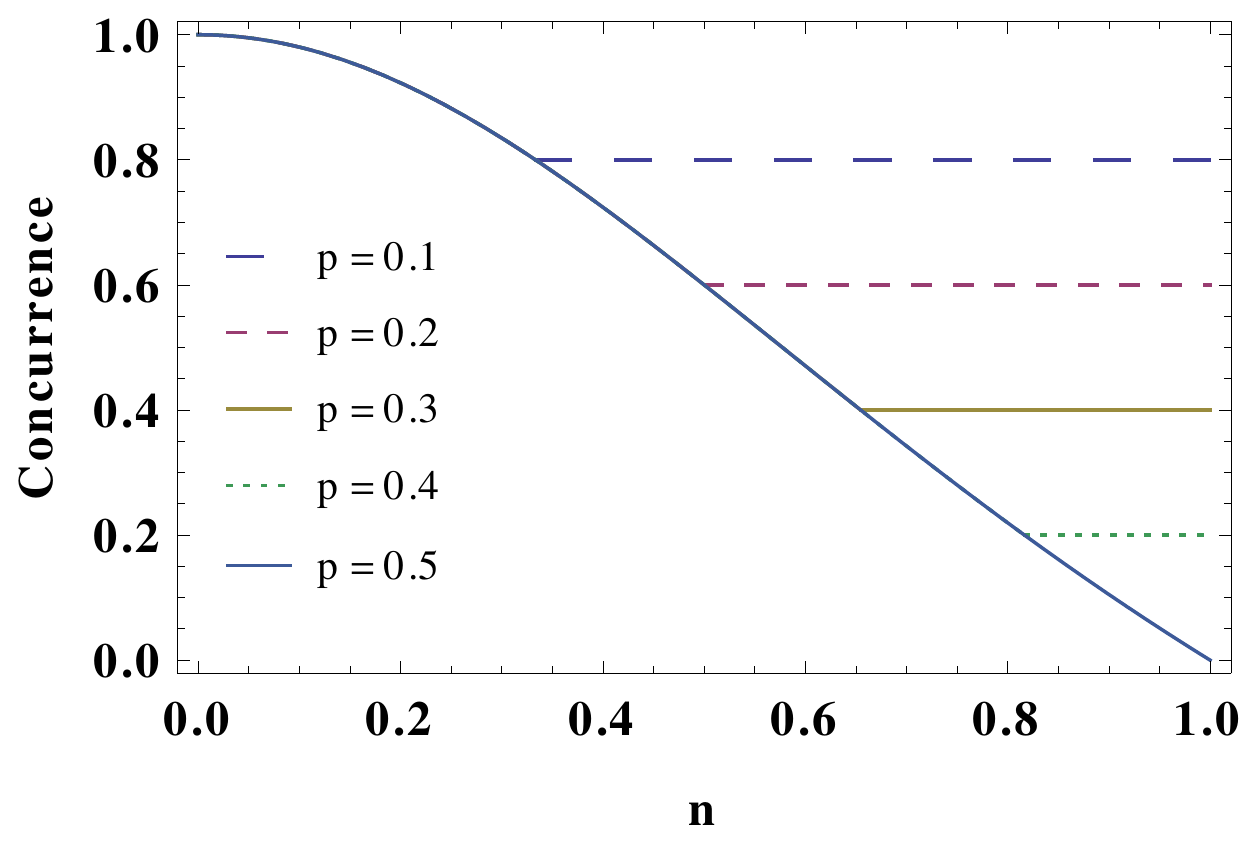}
\caption{Variation of average concurrence with the parameter $n$.}
\label{fig_concurrence}
\end{figure}

 The variation of average fidelity with control parameter $n$ has been displayed in the Fig.(\ref{fig_fidelity}).
It is interesting to point out that Charlie has full control over the protocol when the state is GHZ, {\it i.e.}, $p=0.5$. However, for other values of $p$ and some range of $n$, Charlie does not have any control as average fidelity remains constant. To understand this behavior we compute concurrence \cite{Wooters} of the states $\ket{\psi_+}_{AB}$ and $\ket{\psi_-}_{AB}$ and then average it to find out the average concurrence as,
\begin{equation}
\mathcal{C}_{\mbox{av}}=p_+\mathcal{C}_++p_-\mathcal{C}_-.
\end{equation}
We plot the average concurrence with control parameter $n$ for some values of $p$. The nature of Fig. \ref{fig_concurrence} is exactly same as the nature of Fig. \ref{fig_fidelity}.

The reason is that for a pure state fidelity is related to the concurrence by the formula $\mathcal{F}=\frac{2}{3}(1+\mathcal{C})$. In our case, average fidelity is $\mathcal{F}_{\mbox{av}}=p_+\mathcal{F}_++p_-\mathcal{F}_-=\frac{2}{3}(1+p_+\mathcal{C}_++p_-\mathcal{C}_-)=\frac{2}{3}(1+\mathcal{C}_{\mbox{av}})$.

\section{TMES for Cooperative communication protocols}\label{sec6}

 For multipartite states, there is no unique notion of maximally entangled state like Bell states in two-qubit case. But we can 
 construct a set of states which may be suitable for a particular protocol and those states that can execute the protocol
  maximally. We have seen that apart from the GHZ state, the state (\ref{threequbit3}) is capable of performing {controlled} QKD
   as well as cooperative teleportation. These resource states are suitable for maximal CoQKD, {\it i.e.} the protocol can be
    carried out for perfect key generation with optimal key rate, as well as maximal cooperative teleportation, {\it i.e.} with unit probability and optimal fidelity. Therefore, these states are  task-oriented maximally entangled states (TMESs) as introduced by the authors in \cite{tmes}.

 In the case of tripartite states, as shown in  \cite{tmes}, one should be able obtain a TMES for teleportation by
  applying a suitable multinary transformation on the
product state of a one-qubit state and a Bell state. We now present these transformations for the resource state.
It would be interesting to see how these states can be realized. A single qubit unitary operation followed by a global unitary on a Bell state would suffice to produce this kind of state:
\begin{equation}
        (U_{12} \otimes \sigma^{0}_{3} )\ket{0}_{1}\ket{\phi^+}_{23}  = \sqrt{p}\ket{0}_{1} \ket{\phi^+}_{23} +\sqrt{1-p}\ket{1}_{1}\ket{\phi^-}_{23},
\end{equation}
where, the operator $U_{12} = \begin{bmatrix}
    \sqrt{p} \sigma^0 & \sqrt{1-p} \sigma^z  \\
    - \sqrt{1-p} \sigma^z  & \sqrt{p} \sigma^0
  \end{bmatrix}$
acts on first two qubits and $\sigma^{0}_{3}$ is  $\sigma^{0}$ acting on the third qubit. $\sigma^0$ is the $2\times 2$ identity matrix and $\sigma^z$ is the Pauli-Z matrix.
\section{Conclusion}\label{sec7}
We have considered the CoQKD protocol for secret key generation in a multi-party
situation, where a secret key is established between two parties with the control or supervision of other parties. The advantage of this protocol is that one or more parties can supervise the secret key generation, thus reducing the chance of dishonesty. For a given multipartite state, it is not
always obvious whether this state can be used for {controlled} QKD or Cooperative teleportation. 
In this paper, we have constructed resource states for maximal {controlled} QKD, {\it i.e.} Co-QKD with perfect key generation and optmal key rate.
These states are also suitable for maximal cooperative teleportation, {\em i.e.}
teleportation with unit probability and unit fidelity. The resource states we have discussed are exhaustive for three-qubit 
case. The efficiency of the protocols depends on the choice of the 
measurement basis by the controlling parties. We 
have explicitly shown the dependence of the key rate of CoQKD protocol and fidelity 
of cooperative teleportation with Charlie's choice of measurement basis. Efficiency 
of the protocol (key rate or fidelity)  is controlled by Charlie. If he chooses a basis set wisely then the 
protocol can be carried out maximally, {\it i.e.} optimal key rate or fidelity. 
In this sense,
the states we have discussed are TMESs. 

For arbitrary choice of basis by Charlie, the collapsed state between Alice and Bob is non-maximally entangled. We have discussed security and key rate in this situation for CoQKD.
Apart from CoQKD, we have shown how to generate
a conference key with the resource state. It turns out that recently introduced Bell inequalities
can be used to determine the security of the conference key protocol. We have also gone 
beyond three-qubit scenario, and
constructed suitable resource states for four-qubit states. We hope that our discussion would
lead to experimental observations of these cooperative schemes.

\acknowledgement
P.A. acknowledges the support from the Department of Science and Technology, India, through the project DST/ICPS/\\QuST/Theme-1/2019.

\section*{Author contribution statement}
Development and construction of the problem have beendone  by  all  the  authors.  A.  Das  and  S.  Nandi  carriedout  most  of  the  calculations.  A.  Das,  S.  Nandi,  andP.  Agrawal  examined  and  checked  the  results.  All  theauthors contributed to the preparation of the manuscript.


\end{document}